\documentclass[preprint,12pt]{elsarticle}

\usepackage{hyperref}
\usepackage{amsmath}
\usepackage{amssymb}
\usepackage{graphicx}
\usepackage{color} 
\usepackage{subfigure}
\usepackage{url}

\journal{Neural Networks}

\biboptions{authoryear}

\begin{document}

\begin{frontmatter}

\title{Neural coordination can be enhanced by occasional interruption of normal firing patterns: A self-optimizing spiking neural network model}

\author[address1]{Alexander Woodward\corref{mycorrespondingauthor}}
\cortext[mycorrespondingauthor]{Corresponding author}
\ead{alex@sacral.c.u-tokyo.ac.jp}

\author[address2,address3]{Tom Froese}

\author[address1]{Takashi Ikegami}

\address[address1]{Graduate School of Arts and Sciences, The University of Tokyo, Komaba, Tokyo 153-8902 Japan}
\address[address2]{Departamento de Ciencias de la Computacion, Instituto de Investigaciones en Matematicas Aplicadas y en Sistemas, Universidad Nacional Autonoma de Mexico, Mexico}
\address[address3]{ Centro de Ciencias de la Complejidad, Universidad Nacional Autonoma de Mexico, Mexico}





\begin{abstract}

The state space of a conventional Hopfield network typically exhibits many different attractors of which only a small subset satisfy constraints between neurons in a globally optimal fashion. It has recently been demonstrated that combining Hebbian learning with occasional alterations of normal neural states avoids this problem by means of self-organized enlargement of the best basins of attraction. However, so far it is not clear to what extent this process of self-optimization is also operative in real brains. Here we demonstrate that it can be transferred to more biologically plausible neural networks by implementing a self-optimizing spiking neural network model. In addition, by using this spiking neural network to emulate a Hopfield network with Hebbian learning, we attempt to make a connection between rate-based and temporal coding based neural systems. Although further work is required to make this model more realistic, it already suggests that the efficacy of the self-optimizing process is independent from the simplifying assumptions of a conventional Hopfield network. We also discuss natural and cultural processes that could be responsible for occasional alteration of neural firing patterns in actual brains.

\end{abstract}


\end{frontmatter}


\section{Introduction}


The class of recurrent Hopfield neural networks, first described by Hopfield (\citeyear{Hopfield_1982}), has traditionally been employed for two distinct kinds of tasks. On the one hand, these networks can be trained to form an associative memory of neural activity patterns. Neural activity is set to the activation pattern to be memorized and Hebbian learning is applied in order to turn that pattern into an attractor. One drawback is that spurious attractors are easily formed and these do not represent any target pattern. On the other hand, Hopfield networks can also be used to find solutions to constraint satisfaction problems \citep{Hopfield_1985}. The connection weights are set to represent the constraints between the components of the target problem, the network's activity is initialized to some starting configuration, and the activity is then allowed to converge to an attractor, which at the same time represents a possible solution to the problem represented by the weights. This process of convergence can be understood as a coordination of component activity so as to satisfy the most constraints given the starting configuration. However, as is well known, the state space of a complex Hopfield network typically exhibits many different attractors of which only a small subset are globally optimal; the rest are local optima that fail to take full advantage of the possibilities of coordination.

Watson and colleagues (\citeyear{Watson_2011,Watson2_2011}) recently discovered that Hopfield networks that combine these two tasks manage to overcome both types of drawbacks. They modified the standard constraint satisfaction procedure by making it iterative and including Hebbian learning. As per usual, the weights of the network are set to represent a specific constraint satisfaction problem. Then they used the following itinerant routine: (1) the neural network activity is initialized to a random configuration, (2) the activity is allowed to converge to an attractor, and (3) after this point a small amount of Hebbian learning is applied. What is the effect of repeating this three-step procedure? As might be expected, the neural network forms an associative memory. However, in this case it is not a memory of external patterns, but rather of the different attractor configurations that the network has visited. Over time the network will thereby reconfigure its weight space until most (if not all) initial activity configurations lead to the same attractor, which happens to be one of the best solutions to the original constraint satisfaction problem.

Two properties of Hopfield neural networks are responsible for this useful self-optimization process. First, it has been proven that there is a positive correlation between the width (localizability) and depth (optimality) of a basin of attraction \citep{Kryzhanovsky_2008}, which means that better constraint solutions are visited comparatively more frequently and are therefore reinforced more often. Second, the self-optimization process takes advantage of the learning neural network's ability to generalize over the training set, i.e. the visited attractors. In this case the reinforcement of spurious attractors is actually desirable. For as long as the problem in weight space is decomposable in some manner, the reinforcement of a visited attractor at the same time reinforces other attractors that are partially composed of similar configurations - even if the network has not previously encountered them after one of the re-initializations. In this manner the basins of attraction of still unvisited global optima will become enlarged, and therefore more easily found, even if they are normally extremely difficult to locate. This is effectively the same as if the neural network is translating its original constraint optimization problem into a higher-dimensional organizational space to make it easier to solve, but without making use of any a priori knowledge of the problem domain \citep{Watson3_2011}.

Given the generality of the mathematics underlying the Hopfield network, which is isomorphic to the famous class of Ising models in statistical mechanics \citep[Ch.~13]{Rojas_1996}, as well as the simplicity of the self-optimization process, we can expect this process to govern the emergence of coordination in a wide range of systems. Even the need for true Hebbian learning can be relaxed. In the case of social systems it has been shown that habituation of the behaviors that constitute attractor configurations is sufficient to realize a similar structural self-optimization process \citep{Davies_2011}. Nevertheless, it remains to be verified that the process proposed by Watson and colleagues remains effective when it is implemented in more biologically realistic neural networks, and to provide an interpretation of the necessary periodic deviations from converged behavior (i.e., step 1). 


In this work we created a spiking neural network model that emulates the properties of a traditional Hopfield network with saturated linear transfer (rather than binary threshold) functions, and with real-valued (rather than integer) weights. Our main aim was to demonstrate that the combination of Hebbian learning with occasional alteration of normal neural network activity also leads to the emergence of global neural coordination in such a spiking neural network. Although further modeling work is required to confirm that this self-optimizing process can be operative in even more realistic neural networks, here we managed to show that it is independent of the simplifying assumptions of the conventional Hopfield network. We only interpret the spiking network as a Hopfield network in order to demonstrate that self-optimization is indeed taking place in an equivalent manner.

Additionally, Hebbian learning is a rate-based learning method and we have proposed a form of Hebbian learning in a timing based system, by using heterosynaptic plasticity \citep{Bailey_2000,Huang_2004} and spike-timing dependent plasticity (STDP), equivalent to that found in traditional Hopfield networks, thus making a connection between rate-based and temporal neural encoding systems. Temporal coding based systems have a number of advantages such as not being affected by synaptic depression and being able to achieve a high rate of computation at biological realistic firing rates \cite{Maass_1997b}. In the same paper, Mass et al. have pointed out that from recent experiments `it is in fact questionable whether biological neural systems are able to carry out analogue computation with analogue variables represented as firing rates' (p. 355). 

The rest of this article unfolds as follows. First, we discuss our methods in more detail, paying special attention to how we applied the ideas gained from studies with Hopfield neural networks to the spiking neural network model. Second, we present the results of our investigation, which demonstrate that Hopfield dynamics and the process of self-optimization of neural coordination identified by Watson and colleagues is also effective in spiking neural networks. Finally, we evaluate the plausibility of the spiking neural network model when compared to real nervous systems. We also briefly discuss what could be the natural and cultural causes of occasional alteration of normal neural activity in human brains.

\section{Methods}
 
The large number of recurrent connections in the brain seem to be the mechanism behind its associative memory capabilities. A well known computational neural network architecture that also exhibits such properties is the Hopfield network, first described in \cite{Hopfield_1982}, and with graded neuron response in \cite{Hopfield_1984}. In the following we show how it is possible to transfer some of the key concepts of the Hopfield network to a biologically more realistic spiking neural network.


In early versions of the Hopfield network \citep{Hopfield_1982} the binary state of a neuron was taken to abstractly represent whether that neuron was not firing (0) or firing at maximum rate (1). In later versions \citep{Hopfield_1985}, as well as in more recent elaborations of this kind of network architecture such as including negative self-feedback \citep{Nozawa_1992}, or the continuous-time recurrent neural network \citep{Beer_1995}, a nonlinear function of a neuron's state is typically interpreted to be its mean firing rate. 

It is therefore reasonable to assume that, conversely, the firing rate of a spiking neuron can be interpreted as the state of a traditional Hopfield neuron. However, as alluded to in the introduction, doubt has been raised as to whether firing rates in a biological neural system would be even appropriate for carrying out Hopfield network dynamics. Since temporal coding systems have the potential to achieve a high rate of computation at biological realistic firing rates we therefore describe a way to implement Hopfield network dynamics using temporal coding in a spiking neural network. 

To measure constraint satisfaction by means of neural coordination, the spike-timing encoding of analog values can then be inserted into the Hopfield energy function where attractor states of the network correspond to local minima in the energy landscape (by convention, larger negative values in the Hopfield energy function are more optimal). Our proposed heterosynaptic Hebbian learning approach, described in Section \ref{sec:hetero}, can then be used to update weights in the spiking network as part of the self-optimization process. 

 

\subsection{The Hopfield network}

A Hopfield network is a fully interconnected neural network, usually with symmetric connections between nodes, where $H$ represents the network state: $H = \langle s_{1},\dotsc s_{n} \rangle  \in [0,1]^{n}$. 
 
Hopfield dynamics can be described through updates to neuron states; for the $i$th neuron $s_{i}$:
\begin{equation}
s_{i}(t+1) = \theta \left[ \sum_{j}^N \omega_{ij} s_{j}(t) \right]
\label{equ:hopfield}
\end{equation}

where $\omega_{ij}$ is the weight between neurons $i$ and $j$, collectively described by $\Omega = \langle \omega_{1},\dotsc \omega_{n} \rangle  \in [-1,1]^{n}$ and $\theta$ is a transfer function, here we use a saturated linear transfer function. At each time step a single neuron can be randomly chosen for update, or all neurons can be updated in parallel, the dynamical properties remain either way. Certain network configurations can be imprinted into the network in a one-shot manner as follows:
$ 
\Delta w_{ij} = \rho \sum_{s=1}^{M} (2V_{i}^{s}-1)(2V_{j}^{s}-1)
\label{eqHopfield}
$, where $V$ is a normalized input value, $\rho$ is an input scale factor, and $M$ is the number of configurations to imprint. 

An energy function is associated with the Hopfield network as follows:
\begin{equation}
E (\Omega(t)) = -\sum_{ij}^N \omega_{ij} s_{i}(t) s_{j}(t)
\end{equation}

where smaller values are considered more optimal. Given a symmetric weight matrix this function is guaranteed to be monotonically decreasing until it reaches a fixed-point attractor. Under typical conditions this will be an equilibrium state that only partially satisfies the constraints between neurons, since equilibriums representing globally optimal coordination are comparatively rare.

\subsection{Self-optimization process}

Repeatedly restarting the Hopfield network from random configurations allows the network to settle into a number of local attractors. Attractors with lower energies are visited more often than those with higher energies due to their larger basins of attraction. For networks with symmetric weight matrices it has been found that attractor basin depths are positively related to their width \citep{Kryzhanovsky_2008}. Self-optimization involves imprinting these visited attractor states into the network weights. This process enlarges the basins of attraction of better minima, even if they have not yet been visited (as long as they are partially similar to visited attractors), and eventually allows the restructured neural system to consistently settle into a more globally optimal equilibrium from any starting configuration of neural activities. This process was originally described in \cite{Watson_2011}.

More specifically, the network is repeatedly perturbed into a new random state configuration and allowed to relax from this state into a new attractor - here, relaxation refers to the period during which the network settles into an attractor from an arbitrary initial configuration. Each of these relaxations lasts for $\hat t$ time steps and the network configuration is imprinted using the following Hebbian learning process as defined in \cite{,Watson_2011}:

\begin{equation}
\label{equ:learning}
\omega_{ij}(t+1) = \omega_{ij}(t) + \delta s_{i}(t)  s_{j}(t)
\end{equation}

for all $w_{ij}, i \neq j$, where $\delta>0$ is a learning rate constant, and the weights $\Omega$, were restricted to be between $[-1,1]^{n}$. Learning could be done at every time-step but for algorithmic considerations this was performed at the end of each relaxation period. As long as it is assumed that the network spends most of its time in the attractor the overall effect is the same. Typically, this iterative process continues until a single global attractor remains (i.e. the system reaches the same energy state from any initial configuration).

The original energy $ E_{S}^{0}$ can obtained by using the current network states along with the networks weights before modification: 
\begin{equation}
\label{equ:orig_energy}
E^{0}(\Omega(t=0)) \equiv -\sum_{ij}^N \alpha_{ij} s_{i}(t) s_{j}(t)
\end{equation}

where $a_{ij} \equiv \omega_{ij}(t=0)$. In other words, we use the original weights $a_{ij}$ along with the neuron values $s_{i}$ from the modified network to evaluate $E$. This allows us to compare the energy level on which the network finally converges with the levels that would have been obtained in the original system. In other words, we can verify whether the current state vector is a good solution to the original constraint satisfaction problem.

From \cite{Watson_2011}, conditions on the system are that 
1) dynamics of the system exhibit multiple attractors,
2) the system configurations are repeatedly relaxed from different random initial conditions such that the system samples many different attractors on a timescale where connections change slowly, and
3) the system spends most of its time at attractors.

Additional requirements are that the learning rate must be small so that a wide number of attractors are visited and poor local minima are not reinforced. Also, the time for the system to reach an attractor must be less than the relaxation period $\hat t$ so that the network can consistently reach stable attractor states.

The benefits of this self-optimization process is that it can allow a system to reliably find low-energy configurations - and sometimes more quickly - and that these properties are realized spontaneously without any a priori knowledge of the problem domain.

\subsection{Emulating a Hopfield network using a spiking neuron model}

We want to implement the same self-optimization process, as described in the previous section, in a spiking neural network model. In the following section we therefore introduce the spiking neuron model, we then explain how to interpret its spike timings as encoding a Hopfield neural state, and we describe the spiking network's learning algorithm. As will become evident, the overall self-optimization process remains the same as before.

\subsubsection{Leaky integrate and fire neuron model}

The basic leaky integrate and fire model, as described in \cite{Gerstner_2002}, is as follows:
\begin{equation}
\tau_{m} \frac{\mathrm d u}{\mathrm d t} = - u(t) + R I(t)
\end{equation}

where $u(t)$ is the membrane potential, $\tau_{m}$ is the membrane time constant $I(t)$ is a time varying input current from presynaptic neurons. It can be thought of as operating as a parallel $RC$ circuit, where $R$ is a resistor through which a current can pass through.  

Here the firing event $t^{(f)}$ is given by threshold:

\begin{equation}
t^{(f)}: u(t^{(f)}) \geq v 
\end{equation}

where $v$ is a firing threshold. After a firing event the membrane potential is reset to a low value ($0$ was used in our simulations). All neurons go into a refractory state for a time $\tau_{ref}$. The firing event causes an electrical spike which travels along a synapse and is received by a connected neuron, after some transmission delay time, $\tau_{delay}$ (in this work the same value was used for all neurons). The effect of a spike train at a receiving neuron can be modeled as the summation of postsynaptic currents, which change the membrane potential of a neuron (postsynaptic potential). Thus, the stimulation from synaptic currents for a neuron $i$ is formed from 

\begin{equation}
I_{i}(t) =  \sum_{j}  \omega_{ij}  \sum_{f} \beta(t-t_{ij}^{(f)} - \tau_{delay})
\end{equation}

where $t_{ij}^{(f)}$ represents the $f$th firing event time of neuron $j$ along the connection between $i$ and $j$. The weight $\omega$, as defined in Eq. \ref{equ:hopfield}, determines whether $\beta$ results in an excitatory or inhibitory postsynaptic potential. In our model a unit pulse was used to model the synaptic current pulse: 
\begin{equation}
\beta(t) = \left \{  \begin{array}{rcl}
0 & if & t < 0 \\ 
1 & if & 0 \leq t \leq 1 \\\ 
0 & if & t > 1 \\\end{array} \right .
\end{equation}
 
whose width was then scaled proportional to the time step $\Delta t$ and keeping a unit integral. In reality an approximation to the unit pulse could naturally arise from a collective burst of neural activity since the summation of differently timed pulses could model a continuous function \citep{Maass_1997b}.

\subsubsection{Emulating Hopfield analog values with spike time encoding}

The Hopfield network is a rate-based system where neurons have analog values representing mean firing rates. In this section we describe a method to emulate such a system using spike-timings to encode analog values. Analog computation and Hopfield dynamics in spiking neural networks was first described by \cite{Maass_1997a}, and further developed in works such as \cite{Bohte_2002} and \cite{Tino_2005}. Spike-timing based neural networks have also been implemented in Complementary Metal-Oxide Semiconductor (CMOS) hardware, an integrated circuit technology, in works such as \cite{Tanaka_2009}. Consider a Hopfield network with real-valued, graded neuron response, where $H$ represents the network configuration: $H = \langle s_{1},\dotsc s_{n} \rangle  \in [0,1]^{n}$. 

\begin{figure}[h]
\begin{center}
\includegraphics[width=0.7\textwidth, trim=0 20 0 0, clip=true]{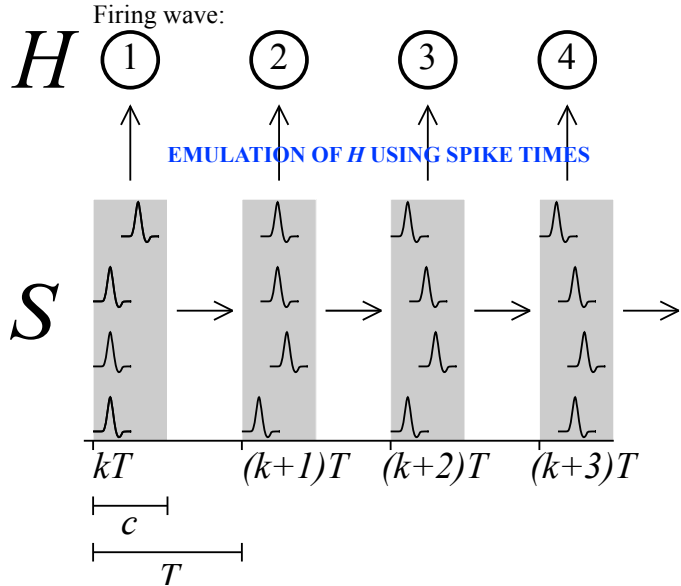}
\end{center}
\caption{Diagram depicting the emulation of a Hopfield network $H$ using the firing times of a spiking neural network $S$ to emulate the values of the Hopfield neurons. The number in the circles indicates $k$, the number of the firing wave, and only four representative spikes are shown per firing wave. Here $T$ represents the period between firing waves and $c$ is the firing window in which neurons may fire. Thus $kT$ is the firing time of firing wave $k$ and so forth.}
\label{fig:coding_diagram}
\end{figure}

The Hopfield state can be emulated by the firing pattern of a spiking neural network $S$, where the $i$th neuron fires at time $kT+c(1-\tilde{s}_{i})$, where $k$ is the $k$th Hopfield state update, $T$ is the period between Hopfield state updates, $c$ is an arbitrary constant representing the window in which all neurons must fire, and the difference between analog values represented in both systems: $|s_{i} - \tilde{s}_{i}|$ `can be made arbitrarily small' \cite[p. 357]{Maass_1997b}. This entire process is shown in Fig.~\ref{fig:coding_diagram}. 

\begin{figure}[h]
\begin{center}
\includegraphics[width=0.4\textwidth, trim=0 30 0 0, clip=true]{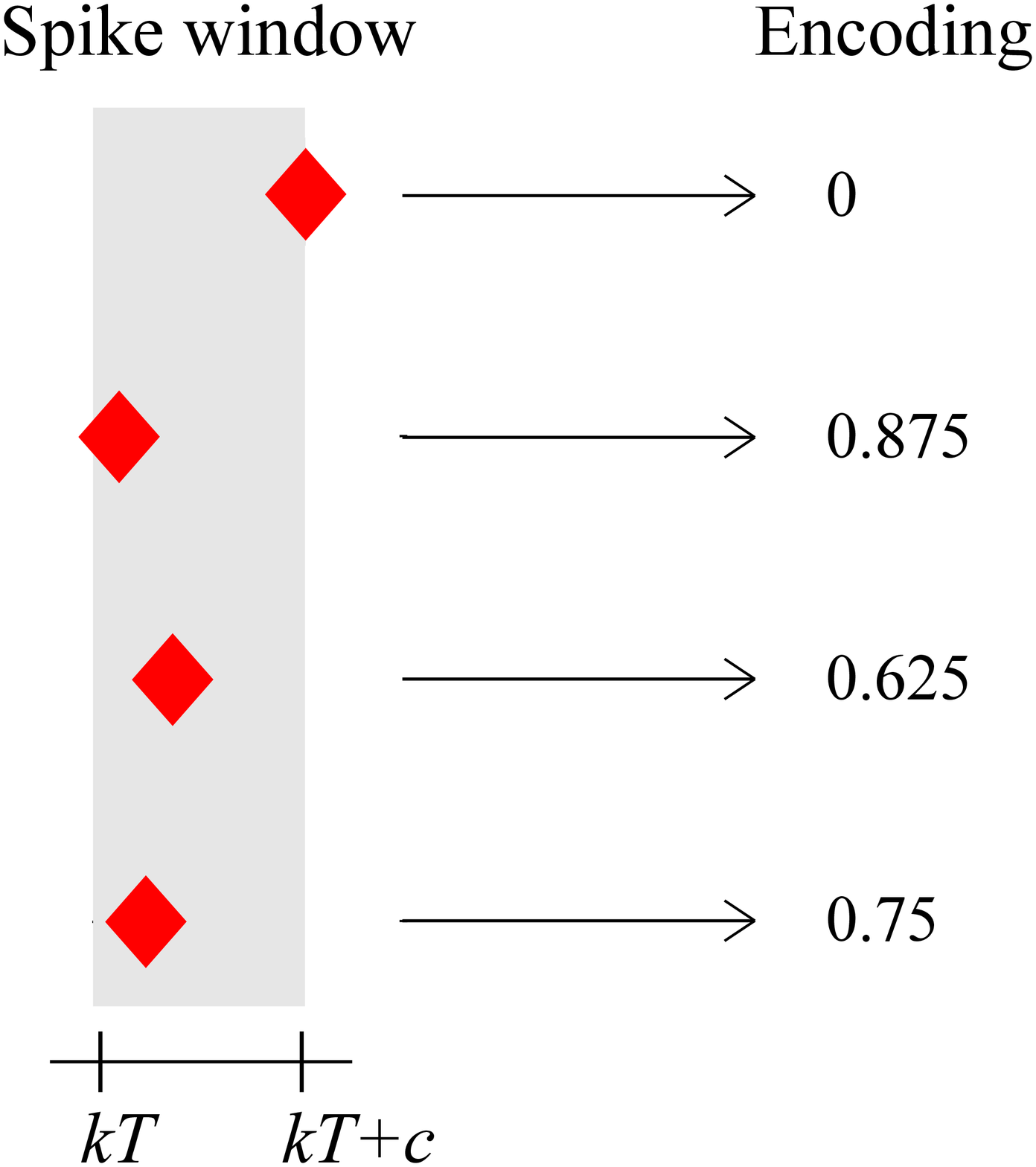}
\end{center}
\caption{The translation of spike times to Hopfield neuron values $s_i$ - for any firing wave shown with a gray background, and with the spikes represented as red diamonds. Only four illustrative spikes are shown. Neurons that spike at the earliest moment $kT$ encode 1, and those that fire at the latest moment $kT+c$ encode 0. Intermediate values are also possible.}
\label{fig:firing_key}
\end{figure}

As an example, a Hopfield neuron with the value $1$ would be encoded in the $k$th firing wave at a time $kT$ and one with the value $0$ would fire at $kT+c$ - this idea is graphically shown in Fig.~\ref{fig:firing_key}.  

The reference timings $kT$, are arbitrarily defined by periodically firing auxiliary neurons and all neurons fire within a time period $[kT,kT+c]$, for simplification simulations used $T= \tau_{delay}$. The first auxiliary neuron spikes at the earliest time $kT$ and all neurons are made to fire at the latest time $kT+c$ through a second auxiliary neural process. Although this method of measuring global neural coordination is somewhat artificial, both of these auxiliary processes are technically necessary in order to define the window $[kT,kT+c]$. At the end of a firing interval the timings can then be read off and measured relative to $kT$ allowing us to interpret them as the analog values of a Hopfield network, which can then be inserted into the energy function Eq.~\ref{equ:orig_energy} to measure the extent of constraint satisfaction.

We emphasize that the two auxiliary neurons are not considered to be a part of the normal functional spiking neural network. They are external central pattern generators that help to subdivide the spiking network's dynamics into distinct periods of meaningful neural activity within which it is possible to interpret each normal neuron's firing timing in terms of Hopfield state analog values within the real-valued interval $[0,1]$.  

Such a neural code, that uses firing patterns of the network, is capable of rapid computation even with slow firing rates, as opposed to a conventional Hopfield rate-based coding approach \citep{Maass_1997b}.

In order to construct a spiking neural network with conventional Hopfield dynamics, namely guaranteed convergence to a fixed-point attractor, we must have a fully interconnected recurrent network and a symmetric weight matrix. Therefore, the weights of normal neurons are initialized to values in the range [-1,1], but with the constraint that weights between any two normal neurons must be symmetric. In addition, all the weights of the two auxiliary neurons are set to +1 and set to fire at kT and kT+c, respectively. 

\subsubsection{Learning in the spiking neural network}
\label{sec:hetero}

We want to preserve the same learning process as in the traditional self-optimizing Hopfield network, described in Eq.~\ref{equ:learning}. Since Hebbian learning is a rate-based process, a plausible mechanism must be defined in order to justify a Hebbian-like process in a spike-timing based system that can work without the causal requirement of straight spike-timing dependent plasticity (STDP). We propose a mechanism using heterosynaptic plasticity and STDP involving a feedback loop. Here the effect of two neurons influences a third neuron (or neuronal cluster) that, assuming feedback connections exist, modulates the connection between the first two. Heterosynaptic plasticity has been explored in works such as \cite{Bailey_2000} and \cite{Huang_2004}. We refer to this type of plasticity as a justification for thinking about Hebbian learning in a spike-timing system that is analogous to the rate-based learning in a traditional Hopfield network. Thus, with a spiking neuron network and its emulation of a Hopfield network formally defined, Eq.~\ref{equ:learning} can be used to modify the weights, through what we call heterosynaptic Hebbian learning. 
\begin{figure}[h]
\begin{center}
\includegraphics[width=0.5\textwidth, trim=0 0 0 0, clip=true]{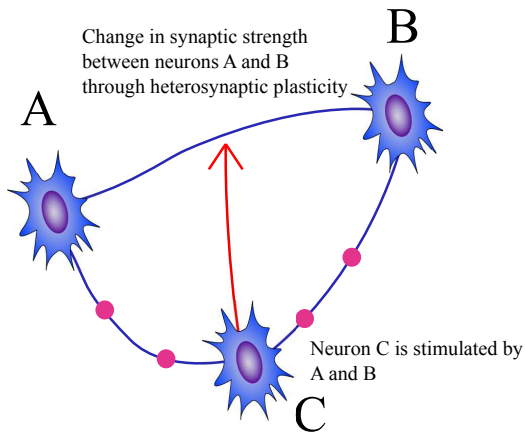}
\end{center}
\caption{Diagram depicting a heterosynaptic learning process. Neuron C fires at $kT$ for a certain wave of firing and is not considered to be within the spiking network that emulates the Hopfield network. Neurons A and B are from that network and are symmetrically connected to each other. Their symmetric connections receive heterosynaptic modification from C. If neurons A and B both spike close to $kT$ they influence C maximally through STDP, which then maximally promotes heterosynaptic modification by C on the connections between A and B. Conversely, modification vanishes when A and B spike at time $kT+c$}
\label{fig:hetero}
\end{figure}

Consider three neurons A, B and C. Neuron C fires at $kT$ for a certain wave of firing and is not considered to be within the functional spiking network that is emulating the Hopfield network (i.e., C is not included in the Hopfield network energy evaluation). Neurons A and B are from the emulating network and are symmetrically connected (as per the definition of network $H$) to each other, and their symmetric connections receive heterosynaptic modification from C. If A and B both spike close to $kT$ they influence C maximally through STDP (i.e. all three neurons spike around the same time $kT$), which then maximally promotes heterosynaptic modification by C on A and B. Conversely, modification of their weights vanishes when A and B spike at time $kT+c$ (i.e. their influence on C in terms of STDP is diminished). Figure~\ref{fig:hetero} depicts this visually.

The non-linear properties afforded by synapses, including their spatial proximity to one another along the dendrite \citep{Koch_2000}, could be capable of approximating the multiplication in Eq.~\ref{equ:learning}, so situations such as when one neuron fires around $kT$ and the other at $kT+c$ have negligible impact, and so on\footnote{For example consider the properties of the logarithmic product: $x\cdot y = log^{-1}(log(x)+log(y))$}.  
   
We realize that there are alternative learning approaches for spike-time coding networks and in future work it would be desirable to try out other more biologically realistic learning algorithms for the spiking network to observe how the self-optimization process is affected. 

\section{Results}

Experiments on the spiking neural network involved a random symmetric weight matrix $\in[-1,1]$ to generate a novel attractor landscape, and a network size of $N=36$ neurons (for computational efficiency - other network sizes, with $N\in[10,100]$, were also tested giving similar results), the learning rate $\Delta = 0.004$, membrane constant $\tau_{m} = 100 \, ms$,a transmission delay time $\tau_{delay} = 20 \, ms$ (for all neurons), for simplicity we also set the period between firing waves $T = 20 \, ms$ and the spiking interval $ c = 10 \, ms$. The continuous time dynamics are evaluated using Euler integration for simplicity in implementation, with update $\Delta t = 0.0125 \, ms$. The relaxation period, $\hat t$ was set at 1000 time steps, giving the system ample time to reach attractor states.

\begin{figure}[h]
\begin{center}
\includegraphics[width=0.9\textwidth, trim=0 20 0 0, clip=true]{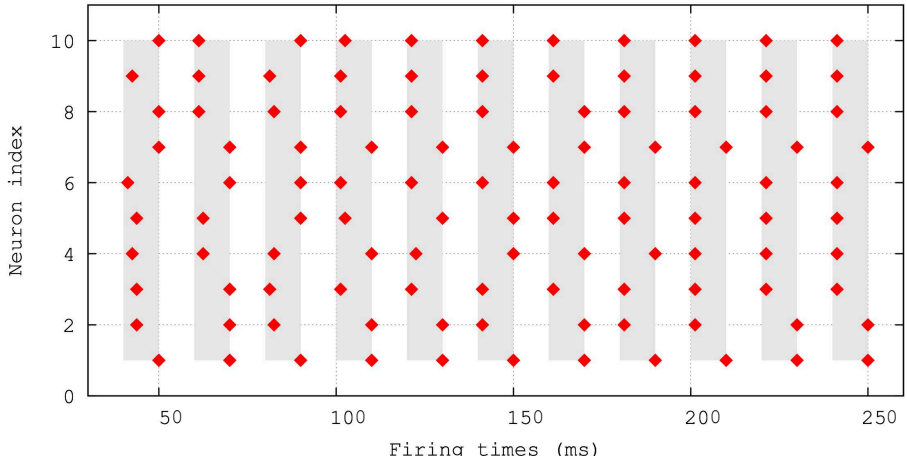}
\end{center}
\caption{Visualization of the firing dynamics for a subset of 10 neurons (for illustrative purposes), showing the spiking neural network (before any self-optimization) settling into an attractor where the firing pattern remains stable after the tenth wave of firing. Shaded regions show the firing interval time window in which a neuron must fire in order to be considered part of that particular firing wave. Summarily, this figure is showing this spiking neural network's typical behavior of convergence to a fixed point of firing dynamics.}
\label{fig:firing_patterns}
\end{figure}

\begin{figure}[]
\subfigure[]{\includegraphics[width=0.5\textwidth, trim=0 20 0 0, clip=true]{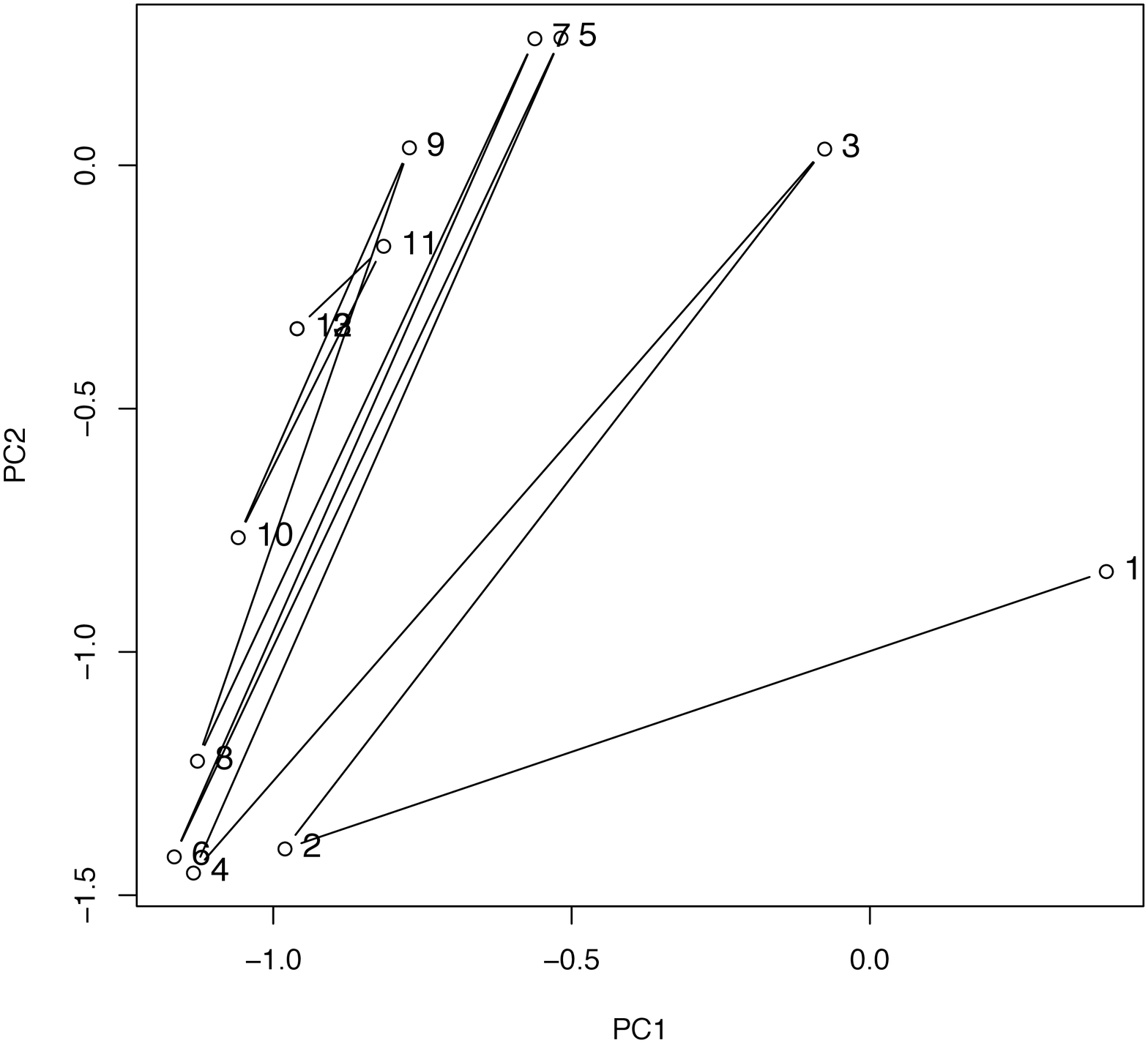}}
\subfigure[]{\includegraphics[width=0.5\textwidth, trim=0 20 0 0, clip=true]{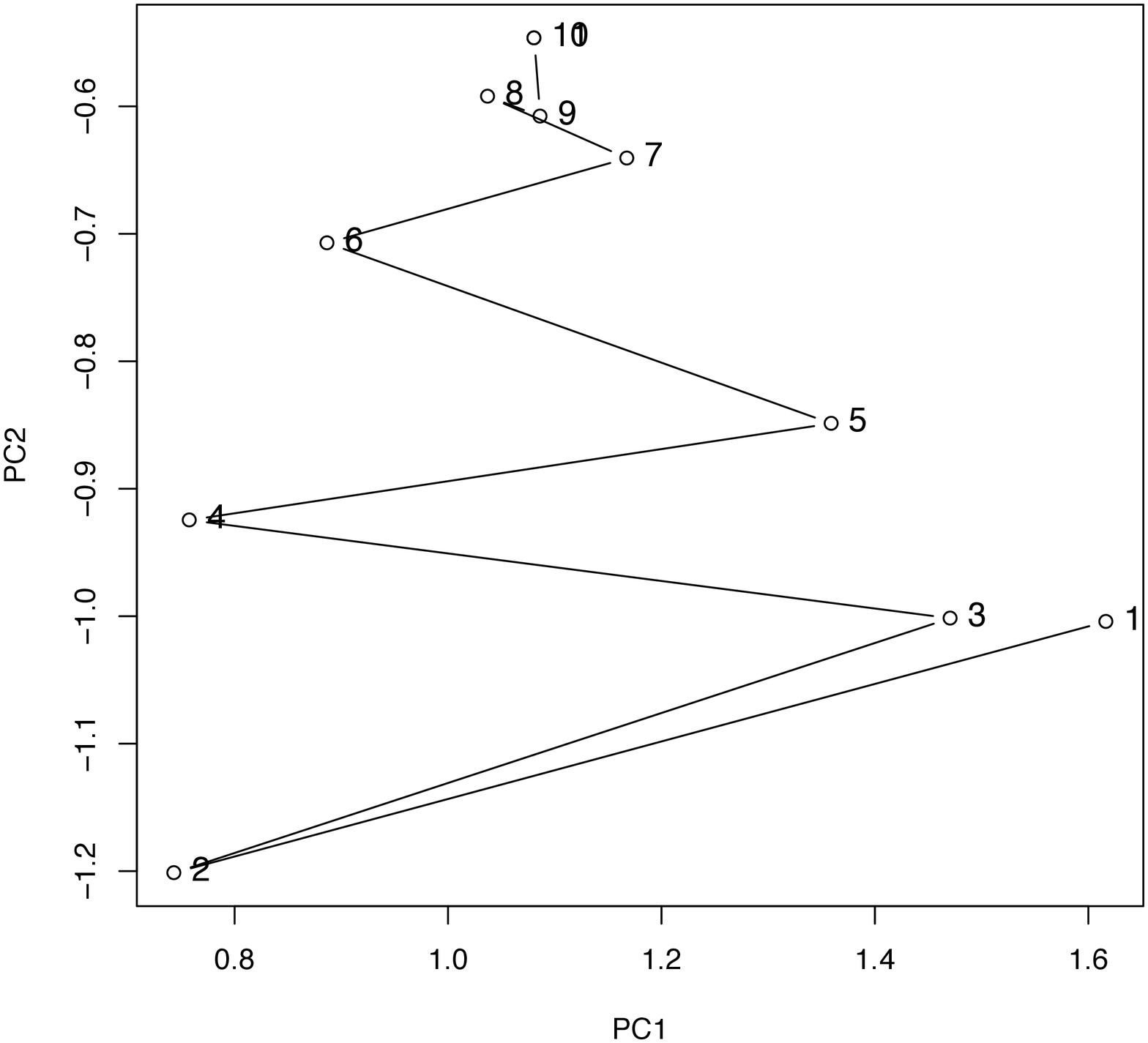}}
\subfigure[]{\includegraphics[width=0.5\textwidth, trim=0 20 0 0, clip=true]{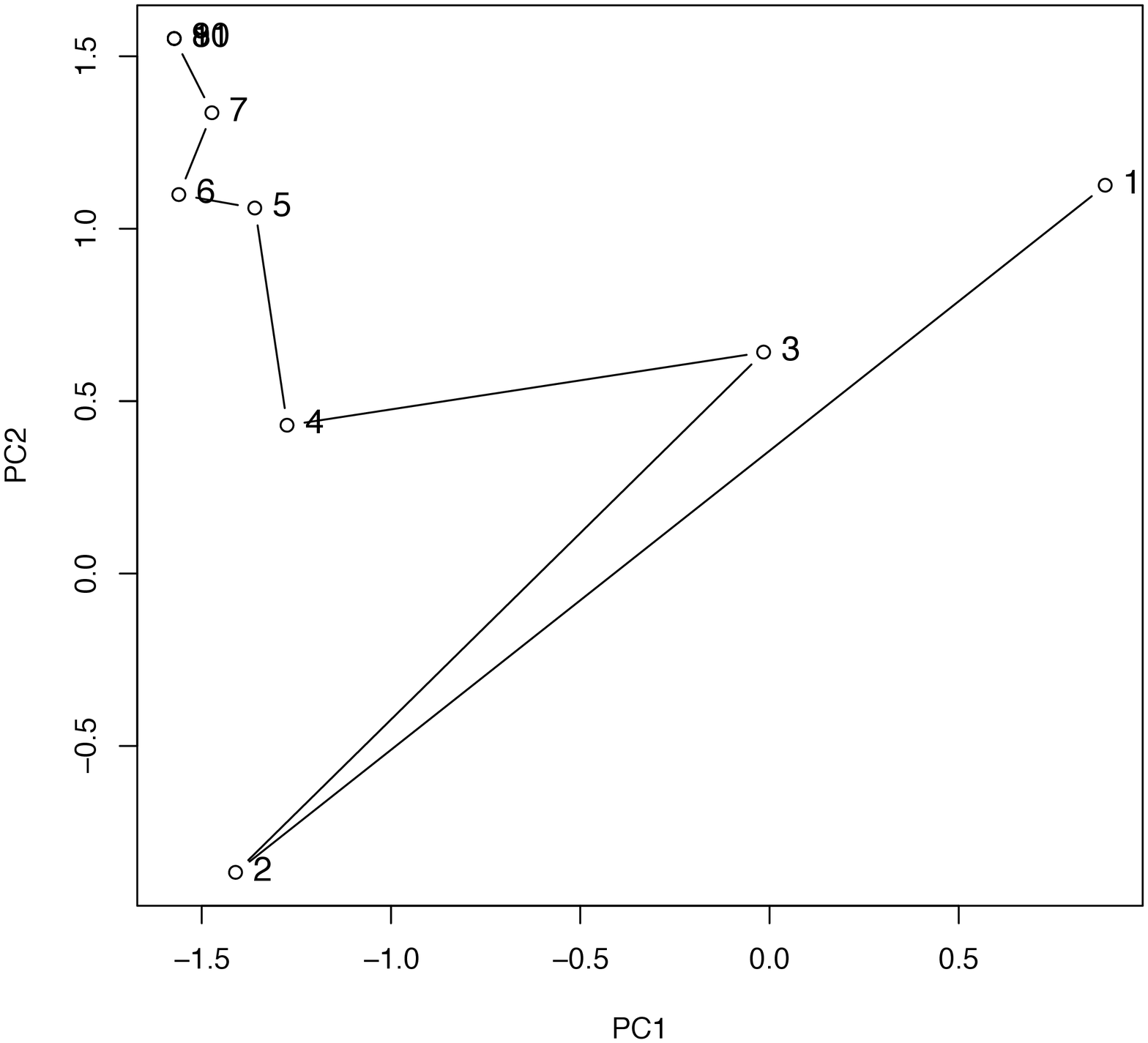}}
\subfigure[]{\includegraphics[width=0.5\textwidth, trim=0 20 0 0, clip=true]{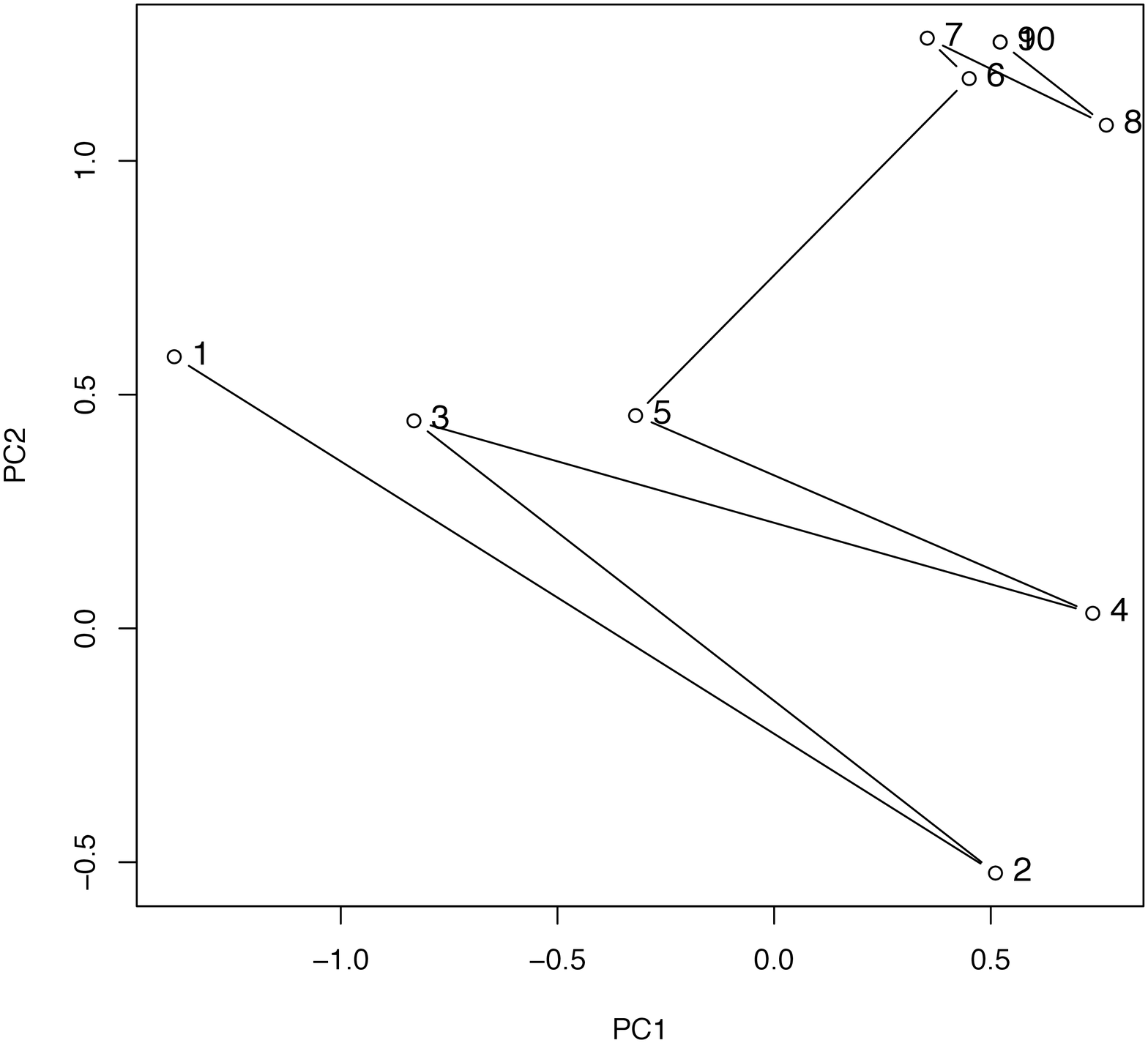}}
\caption{Visualization of attractor dynamics for a number of different initial conditions of the spiking neural network (before any self-optimization). Principal component analysis (PCA) was used to find the axes of maximal variance within the time series data and project the phase space into 2 dimensions. The point labels are in order of firing wave. Trajectories (a)-(d) show the network converging on different attractors, as required by the self-optimization algorithm.}
\label{fig:attractor_vis}
\end{figure}

Hopfield attractor dynamics were successfully implemented, where the system settles into an attractor state in the form of a regular firing pattern at each firing wave (as shown in Fig.~\ref{fig:firing_patterns}). Figure~\ref{fig:attractor_vis} additionally shows attractor dynamics of the network where it converges to different attractors when initialized from different starting conditions. These results show that our spiking neural network implementation may be used for the self-optimization process that creates an associative memory of the different attractors that have been visited.

\begin{figure}[h]
\begin{center}
\includegraphics[width=0.9\textwidth]{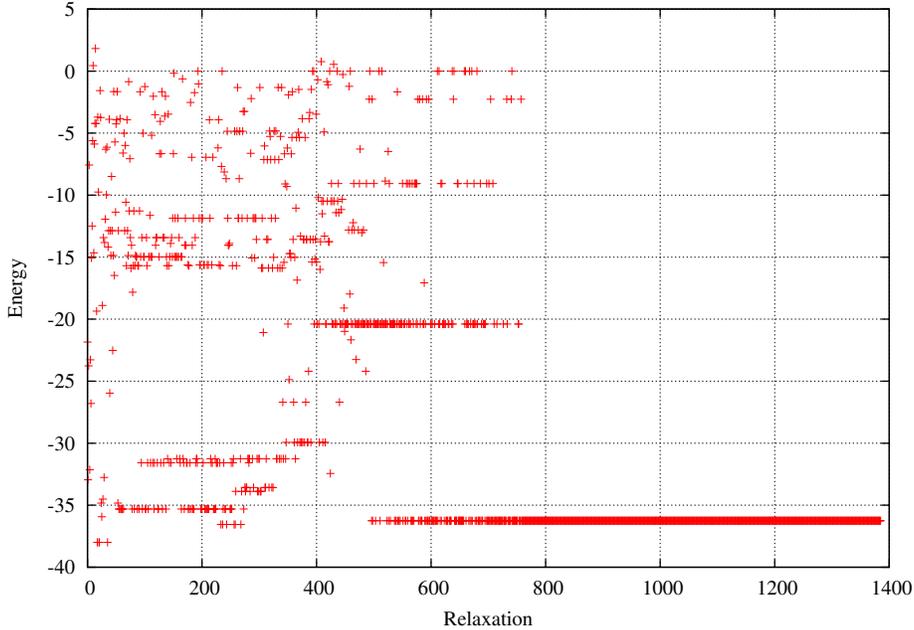}
\end{center}
\caption{An example of the self-optimization process showing the final energy states that the spiking neural network reaches after each relaxation. The energy is measured in terms of the original weight space in order to demonstrate that the process is coordinating a solution to the original constraint problem. After sufficient self-optimization, in this case after approximately 800 relaxations, the network consistently goes to a strong energy minimum from any initial state configuration. A time step of 0.0125 milliseconds was used with 1000 steps per relaxation; after approximately 800 relaxations the system settles into a minimal energy state attractor. }
\label{fig:relaxation_dynamics}
\end{figure}

The structural self-optimizing dynamics are exemplified in Fig.~\ref{fig:relaxation_dynamics}, which shows the final energy states reached by the system from a number of random initial conditions. The self-optimization process is employed and eventually the system consistently reaches a low energy configuration, i.e. an attractor that manages to resolve a significant percentage of constraints between neurons, for the original constraint satisfaction problem from any initial configuration. 

In order to investigate the generality of these results, the network was initialized with another random symmetric weight matrix with values $\in[-1,1]$. This was tested repeatedly for different sets of weight matrices and the self-optimization dynamics was consistently observed showing that such networks are not exceptional - the associative network attractor dynamics (a feature of highly recurrent networks) and Hebbian process are the important properties for self-optimization to be successful. We collected statistics for 100 different networks of size $37$ neurons and found that the average energy level before optimization was -23.10 (with standard deviation of 14.33) and after self-optimization -56.77 (with standard deviation of 0; since the same attractor was always being reached the variance naturally reduced to zero). This improvement in coordination was statistically significant ($p < 0.0001$) due to the decrease in standard deviation. We further verified the generality of this finding by conducting the same test on networks of different sizes and observed the same self-optimizing results. For example: for $N=11$, average energy before optimization: -1.17, standard deviation 2.10, and after self-optimization -7.38; for $N=23$, average energy before optimization: -8.19, standard deviation 7.27, and after self-optimization -28.35 with similar p-values $< 0.0001$. Admittedly, it remains to be seen whether the results could be qualitatively different for spiking neural networks involving up to hundreds and thousands of neurons.

\section{Discussion}

We successfully implemented a spiking neural network that exhibits Hopfield dynamics along with the computational advantages of spike time coding. We were also able to demonstrate that the neural self-optimization process described by \cite{Watson_2011} is independent of the simplifying assumptions of the conventional Hopfield network. The original work by \cite{Watson_2011} used a conventional Hopfield network with a fully interconnected symmetric weight matrix, with weights $\omega_{ij} = {-1,1}$ and binary neurons with discrete activity. Our real-valued spiking neuron model demonstrates that a similar self-optimization process can be found in a biologically more realistic neural network using spike timings. 

We formally measured this self-optimization by interpreting the spiking network as a Hopfield network. This interpretation put certain artificial constraints on the implementation of our spiking neuron model, and future work could develop a less constraining measure of self-optimization. For example, it would be interesting to give a more behavioral interpretation of the optimality of firing patterns by embedding a self-optimizing spiking network in a mobile robot that is required to perform some task. In addition, it would be worthwhile to study what happens when the constraint of symmetric weights is relaxed. Also, we note that each node of the network need not represent the activity of a single neuron, and could abstractly represent clusters of neurons acting as functional sub-networks. Finally, the occasional alteration of firing activity should also be made more realistic.

Indeed, the occasional alteration of normal neural activity, which we simulated as a total randomization of the network's state space, is an essential aspect of the self-optimization process. The more each `reset' state deviates from previously visited state configurations, the more likely it is that the neural network will converge on a novel attractor, and thereby implicitly learn more about the layout of its own overall state space. The variety of visited local optima provides the basis for the network's ability to generalize, allowing it to spontaneously converge on forms of neural coordination that optimally resolve global constraints between neural activity. However, despite the fact that this occasional alteration is fundamental for the efficacy of the self-optimizing mechanism, Watson and colleagues have not discussed what kind of process could realize this alteration in biological neural networks. Here we would like to mention some possible neural mechanisms that deserve to be studied more closely in future work listed in order of increasing temporal scale. 

\begin{itemize}
\item The emergence and dissolution of cell assemblies: \cite{Varela_1999} defines neuronal or cell assemblies (CA) as a distributed subset of neurons with strong reciprocal connections (much like the structure of a Hopfield network), whose coherence of activity integrates more basic neural events. The CA is perhaps the closest biological match for the spiking neural network we have presented in this paper. What is interesting here is that a CA must have a time of emergence within which it arises, stabilizes, and dissolves again, only to begin another cycle. Through the self-optimization process our spiking neural network tends to go to an optimal attractor, which may correspond to a functionally efficacious CA in the sense of Varela. Importantly, a spike-timing based system could operate at a short time-scale appropriate for such processes, whereas a traditional rate-based formulation of a Hopfield network would operate at too long a time-scale.  

\item The sleep-wake cycle: It would also be worthwhile to investigate whether one function of circadian rhythms could be to provide regular resets from normal neural functioning. We note that if sleep could perform the resets needed for neural self-optimization, then we should expect to find evidence that sleep dysfunction is related to non-optimal mental conditions. In this regard it is interesting that schizophrenia is increasingly becoming associated with sleep dysfunction, including abnormality of circadian rhythm and reduced sleep spindles \citep{Wilson_2012}. As a metaphor, we do not think that our current neural network model has sleep or awake phases, or that brain activity during sleep would be a randomization of neural activity as in our current work. But the reset within our system facilitates the exploration of a wide number of memory states (attractors) and this is the important point; sleep seems to involve such an exploration of unusual states.

\item The ritual cycle: It is a general feature of societies that some of their cultural practices temporarily modify the state of consciousness of participating individuals \citep{VanGennep_1908,Turner_1969}. We hypothesize that this ritualized mind-alteration could serve as a global neural reset. For example, certain altered states involve self-sustaining neural processes that partially decouple normal activity from environmental influence \citep{Froese2_2013}. More specifically, Carhart et al.~(\citeyear{Carhart_2014}) applied an entropy measure to neural states and found that the psychedelic state (here induced by psilocybin) is one of elevated entropy over the normal waking state. Similarly, Muthukumaraswamy et al.~(\citeyear{Muthukumaraswamy_2013}) found that psilocybin desynchronized oscillations in the cortex. Given that high entropy is associated with a high level of randomness, this lends support to our idea that altered states can be approximated as a randomization of neural behavior. Thus, our self-optimization model could help to explain why the application of psychedelic substances in the psychiatric context can lead to improved mental functioning \citep{Kupferschmidt_2014}.
\end{itemize}

In conclusion, this paper shows a connection between rate-based coding systems such as the Hopfield network and temporal based coding systems using neuron spike times. We describe a method to emulate a Hopfield network with Hebbian learning by using a spiking neuron timing based system, showing that a self-optimizing process which increases the basin sizes of more optimal attractors, originally discovered in the context of the conventional Hopfield network by Watson et al. (\citeyear{Watson_2011}), can be replicated with a plastic spiking neural network. While the biological realism of our spiking neural network model needs further improvement, it already provides valuable new impulses for improving our understanding of the functional role of occasional deviations from normal neural activity.

\section*{References}

\bibliography{alexbib,FroeseEndNoteLibrary}

\end{document}